\documentclass[preprint,showpacs]{revtex4}

\usepackage{amsmath}

\usepackage{graphics}
\usepackage{graphicx}

\DeclareMathOperator*{\SumInt}{%
\mathchoice%
  {\ooalign{$\displaystyle\sum$\cr\hidewidth$\displaystyle\int$\hidewidth\cr}}
  {\ooalign{\raisebox{.14\height}{\scalebox{.7}{$\textstyle\sum$}}\cr\hidewidth$\textstyle\int$\hidewidth\cr}}
  {\ooalign{\raisebox{.2\height}{\scalebox{.6}{$\scriptstyle\sum$}}\cr$\scriptstyle\int$\cr}}
  {\ooalign{\raisebox{.2\height}{\scalebox{.6}{$\scriptstyle\sum$}}\cr$\scriptstyle\int$\cr}}
}

\def\m@thcombine#1#2{%
  \setbox0=\hbox{$#1$}
  \setbox1=\hbox{$#2$}
  \ifdim\wd0>\wd1
    \setbox0=\hbox to\wd1{\hss\box0\hss}
  \else
    \setbox1=\hbox to\wd0{\hss\box1\hss}
  \fi
  \mathop{\vcenter{
    \offinterlineskip\box0\box1}}}
\def\lesim{\m@thcombine<\sim}
\def\gesim{\m@thcombine>\sim}
\def\lessgtr{\m@thcombine<>}
\def\gtrless{\m@thcombine><}
\newcommand{\bra}[1]{\left\langle #1 \right|}
\newcommand{\ket}[1]{\left| #1 \right\rangle}
\newcommand{\braket}[3]{\left\langle #1 | #2 | #3 \right\rangle}
\newcommand{\rbra}[1]{\left( #1 \right|}
\newcommand{\rket}[1]{\left| #1 \right)}

\newcommand{\Vhat}{\hat{V}}
\newcommand{\Dhat}{\hat{D}}
\newcommand{\That}{\hat{T}}
\newcommand{\hhat}{\hat{h}}

\newcommand{\calHz}{{\cal{H}}_0}
\newcommand{\vecr}{\mbox{\boldmath $r$}}
\newcommand{\vecrs}{\mbox{\boldmath $r$}\sigma}
\newcommand{\vecrst}{\mbox{\boldmath $r$}\tilde{\sigma}}
\newcommand{\vecrsp}{\mbox{\boldmath $r$}'\sigma'}

\newcommand{\vphi}{\varphi}

\newcommand{\phibar}{\overline{\phi}}
\newcommand{\phibari}{\overline{\phi}_{\tilde{i}}}

\newcommand{\phicari}{\overline{\phi}_{{i}}}
\newcommand{\phicarj}{\overline{\phi}_{{j}}}
\newcommand{\phicarn}{\overline{\phi}_{{n}}}
\newcommand{\phicarp}{\overline{\phi}_{{p}}}

\newcommand{\calV}{{\cal{V}}}

\newcommand{\calA}{{\cal{A}}}

\newcommand{\calY}{{\cal{Y}}}
\newcommand{\del}{\partial}
\newcommand{\eps}{\epsilon}
\newcommand{\Tr}{{\rm Tr}}
\newcommand{\rhohat}{\hat{\rho}}

\begin{document}

\title{Continuum quasiparticle random phase approximation for astrophysical
direct neutron capture reaction of neutron-rich nuclei}

\author{Masayuki Matsuo}

\address{
Department of Physics, Faculty of Science,  Niigata University,
Niigata 950-2181, Japan }

\date{\today}

\begin{abstract}
We formulate a many-body theory to calculate the cross section of direct radiative neutron capture reaction
by means of the 
Hartree-Fock-Bogoliubov mean-field model and the continuum quasiparticle
random phase approximation (QRPA).  A focus is put on   
very neutron-rich nuclei and low-energy neutron
kinetic energy 
in the range of O(1 keV) - O(1 MeV), relevant  for 
the rapid neutron-capture process of nucleosynthesis. We begin with 
the photo-absorption
cross section and the E1 strength function, then,  in order to
apply the reciprocity theorem, we  
decompose the cross section  into partial cross sections corresponding to different 
channels of one- and two-neutron emission decays of photo-excited states.
Numerical example is shown for the photo-absorption of $^{142}$Sn
and the neutron capture of $^{141}$Sn.
\end{abstract}


\maketitle

\section{Introduction}\label{sec:intro}

The radiative neutron capture, i.e. $(n,\gamma)$ reaction, is one of the fundamental nuclear
reactions essential in various nucleosynthesis models. 
In the rapid neutron-capture process (r-process), relevant
to the origin of heavy elements,  the reaction takes place 
in short-lived neutron-rich nuclei, for which direct experimental
measurement of the neutron capture cross section is
practically impossible. Naturally, an alternative method to measure 
the inverse reaction, e.g. the Coulomb dissociation, has been
considered, but the actual application is quite limited at present
even though the experimental possibilities increase with the advances of
the RI beam facilities, 
see for example
\cite{Sasaqui2005,Nakamura-19C}.

The neutron capture reaction is often
classified into two different processes depending on the neutron separation energy
or the excitation energy after the capture (see, e.g., \cite{Arnould07} as a review). 
One is the compound process in which 
formation of compound states after absorption of a
neutron is assumed, and statistical models are often
employed. The compound process has been adopted to describe the slow and rapid
neutron-capture processes 
which take place in stable nuclei  or unstable nuclei with large neutron separation energy
sufficient to give excitation energy to form the compound states. The main building
blocks of the model is the neutron transmission coefficient for the formation of
compound states 
and the gamma-decay strength function for the statistical gamma decay. 
Recently new modes
of dipole excitation such as the pygmy resonance and the soft dipole 
excitation\cite{Hansen87,Suzuki90,Esbensen,Paar07} have 
attracted attentions
since influence of the new modes on
the r-process nucleosynthesis was pointed out
\cite{Goriely98}.
 Motivated
with this possibility, microscopic many-body models
of electric multipole responses, developed on the basis of the density functional theories, have been applied to 
the gamma-ray strength function for the compound process calculations
\cite{Goriely-Khan02,Goriely-Khan04,Litvinova09,Avdeenkov11,Daoutidis12,Xu-Goriely14}.

The other is the direct radiative capture process in which a electromagnetic transition
is assumed to take place from an initial state including the incoming neutron 
 to bound final states without forming
compound states. 
It is estimated that  the r-process nucleosynthesis in very neutron-rich nuclei is dominated by the
direct process since the formation of the compound states is not likely
in neutron-rich nuclei due to small neutron separation energy and small
level density at the threshold\cite{Mathews83,Goriely97,Arnould07}. 
Direct neutron capture 
calculations\cite{Lane-Lynn60,Raman85,Mathews83,Mengoni95,Goriely97,Rauscher98, Rauscher10,Chiba08,Xu-Goriely12}
often assumes a simple 
potential picture in which
the initial and final states are described as scattering and bound single-particle states. 
However, one can expect that the new dipole excitation modes may
affect the direct neutron capture process likewise
in the case of the compound process.
To unveil the effect it is necessary to construct
a many-body theory of the direct neutron capture reaction
in which the correlations in multipole modes of excitation 
are  taken into account.
It is the purpose of the present paper to demonstrate
that the
quasiparticle random phase approximation provides such a framework.

The random phase approximation based on the density functional models has been one of the most powerful
theoretical framework to describe  electromagnetic responses of nuclei, including new modes
in exotic nuclei (see, e.g., \cite{Paar07} as a review, and references therein). 
The same is for the photo-absorption cross section. 
Note here that the photo-absorption
reaction may have different final reaction channels if nucleons are allowed to
be emitted from the photo-excited states. Among them, a reaction followed by
one-neutron emission, i.e., $(\gamma,n)$ reaction, is the inverse process of the
relevant $(n,\gamma)$ reaction.  It is then possible to evaluate
the  $(n,\gamma)$ cross section using the reciprocity theorem, provided that
one can calculate 
{\it partial} cross
sections associated with one-neutron emission decays. 
We note here a method of Zangwill and Soven\cite{Zangwill-Soven},
which is used to describe the partial photo-absorption cross section of atoms\cite{Zangwill-Soven} 
and molecules \cite{Nakatsukasa-Yabana} by means of the continuum RPA.
In the case of neutron-rich nuclei, however, the pair correlation plays
important roles\cite{DobHFB1,DobHFB2,Esbensen,Meng-Li,Barranco,Brink-Broglia}, 
and not only one neutron  but also two neutrons can be emitted only with small excitation energy.
We shall show in the present paper that the partial photo-absorption cross sections
corresponding to individual decay channels can be calculated
by applying the Zangwill-Soven method to 
 the continuum
quasiparticle random phase approximation (cQRPA)\cite{Matsuo01,Serizawa09,Mizuyama09}, a version
of QRPA, in which the pair correlation is described with the Bogoliubov theory, and
 the continuum states relevant to the one- and 
two-neutron emission are described with the proper scattering boundary condition.

We remark  that special cares are required to describe capture of a neutron with
very low kinetic energy: the energy range relevant to the r-process is  
$E_n \sim$ O(1 keV) -O(1 MeV), corresponding to the
temperature $T \sim$ O($10^7$) - O($10^9$)  K of possible r-process environments,
and hence we need
a fine energy resolution, which is not required in usual RPA or QRPA descriptions
of nuclear responses.  It should be noted
also that the r-process pass may reach to nuclei close to the neutron drip-line
having very small one-neutron separation energy $S_{1n} \sim 1$ MeV.
In such a case we need wave functions of neutrons up to very 
large distances from the center of the nucleus. Numerical procedures to
meet these requirements are also discussed in the present paper.

\section{Continuum quasiparticle random phase approximation
 for direct neutron capture cross section} 

\subsection{Total photo-absorption cross section in QRPA}

We shall briefly recapitulate the quasiparticle random phase approximation (QRPA)
and its application to a
description of the total photo-absorption cross section in order
to provide a basis for  later discussion.

The photo-absorption reaction is an excitation of a nucleus caused by the electromagnetic
transition. 
Assuming the dominant electric dipole transition (E1 transition) and
the second order perturbation with respect to the photo-nuclear interaction,
the cross section is
given\cite{Ring-Schuck,BM2}
 as
\begin{equation}\label{totphotoabs}
\sigma_\gamma(E_\gamma)=\frac{16\pi^3 e^2 E_\gamma }{3\hbar c} \sum_k |\bra{k}D_0 \ket{0}|^2 
\delta(E_\gamma-\hbar\omega_k)
\end{equation}
with the dipole operator 
$\hat{D}_0=\frac{Z}{A}\sum_p r_pY_{10}(\hat{\vecr}_p) - \frac{N}{A}\sum_n r_nY_{10}(\hat{\vecr}_n)$. Here 
$\ket{0}$  and $\ket{k}$ are the ground and excited states of the nucleus 
with the excitation energy $\hbar\omega_k$. 
The cross section $\sigma_\gamma(E_\gamma)$ is proportional to
the strength function
\begin{equation}\label{strfn0}
S(\hbar\omega)=\sum_k |\bra{k}\Dhat_0\ket{0}|^2 
\delta(\hbar\omega-\hbar\omega_k) 
\end{equation}
with $E_\gamma=\hbar\omega$, 
multiplied with the factor  $f(E_\gamma)={16\pi^3 e^2 E_\gamma /3\hbar c}$.

The strength function is formulated by considering linear response of the system under  an
external one-body field
\begin{equation}
\hat{V}_{{\rm ext}}(t)=\hat{V}_{{\rm ext}}e^{-i\omega t} + \hat{V}_{{\rm ext}}^\dagger e^{-i\omega t}
\end{equation}
with $\Vhat_{{\rm ext}}=\Dhat_0$.
In QRPA, the response
is described on the basis of  the time-dependent Hartree-Fock-Bogoliubov (TDHFB) theory
(which may be called also 
the time-dependent Kohn-Sham-Bogoliubov theory), whose basic equation is 
\begin{equation}
i\hbar\frac{\del}{\del t} \ket{\Phi(t)} = (\hat{h}[R(t)]+\hat{V}_{{\rm ext}}(t))\ket{\Phi(t)}.
\end{equation}
Here  $\ket{\Phi(t)}$ is the time-evolving generalized determinant state
and $\hat{h}[R(t)]=\That + \Vhat_{{\rm mf}}[R(t)]$ is the TDHFB(TDKSB) self-consistent Hamiltonian defined by the variation
of the energy density functional $E_{tot}[R]=\langle \Phi | \That |  \Phi \rangle + E_{x}[R]$ with respect to the generalized density
matrix matrix $R$. 

In the following we assume that the functional is written in terms of quasi-local one-body densities.
The simplest are the one-body density 
$\rho(\vecr)=\sum_\sigma\braket{\Phi}{\psi^\dagger(\vecrs) \psi^\dagger(\vecrs)}{\Phi} $
and the pair-density 
$\tilde{\rho}(\vecr)=\braket{\Phi}{\psi(\vecrst) \psi^\dagger(\vecrs)}{\Phi} $ and
its conjugate $\tilde{\rho}^*(\vecr)$ while
it is not difficult to take into account other quasi-local densities such as
the spin, current, kinetic energy and spin-orbit densities, utilized in the Skyrme
functional models\cite{Bender}. In the following, all these quasi-local densities are denoted as 
\begin{equation}
\rho_\alpha(\vecr)=\braket{\Phi}{\hat{\rho}_\alpha(\vecr)}{\Phi}
\end{equation}
 with
the index $\alpha$ distinguishing the kinds. We also use
 a collective notation $\left\{ \rho \right\}$. Here
$\hat{\rho}_\alpha(\vecr)$ are corresponding one-body operators. 
( In the 
following we assume that the  operators 
satisfy the (anti) hermiticity
$\rhohat_\alpha(\vecr)^\dagger = s_\alpha\rhohat_\alpha(\vecr)$ with $s_\alpha=\pm 1$.)
The TDHFB mean-field $\Vhat_{{\rm mf}}$  is then expressed as
\begin{equation}
\Vhat_{{\rm mf}}[\left\{ \rho(t) \right\}]  =\sum_\alpha  \int d\vecr v_\alpha^{{\rm mf}}(\vecr,t) \rhohat_\alpha(\vecr),
\end{equation}
in terms of  the functional
derivative $v_\alpha^{{\rm mf}}(\vecr,t)= \del E_x[\left\{ \rho(t) \right\}] /\del \rho_\alpha(\vecr)$. 
We also assume that the external field is expressed as
\begin{equation}
\Vhat_{{\rm ext}}  =\sum_\alpha  \int d\vecr v_\alpha^{{\rm ext}}(\vecr) \rhohat_\alpha(\vecr).
\end{equation}

Considering the time-evolution in the linear response approximation, 
we describe the fluctuating part $\ket{\delta \Phi(t)}$ 
of the state vector $\ket{\Phi(t)}$
around the HFB ground state $\ket{\Phi_0}$:
\begin{equation}\label{fluc-state}
(i\hbar \frac{\del}{\del t} - \hat{h}_0)\ket{\delta\Phi(t)}=
(\hat{V}_{{\rm ext}}(t)+ \delta\hat{V}_{{\rm ind}}(t))\ket{\Phi_0},
\end{equation}
where 
\begin{equation}
\delta\hat{V}_{{\rm ind}}(t)= \sum_{\alpha\beta}\int d\vecr 
\kappa_{\alpha\beta}(\vecr)\delta\rho_\beta(\vecr,t)\hat{\rho}_\alpha(\vecr),
\end{equation}
is fluctuation in the TDHFB mean-field $\Vhat_{{\rm mf}}[\left\{ \rho(t) \right\}]$, and 
is  often called the {\it induced field}. It arises from fluctuation in the densities
$\delta\rho_\alpha(\vecr,t)=\braket{\Phi_0}{\hat{\rho}_\alpha(\vecr)}{\delta\Phi(t)}
+\braket{\delta\Phi(t)}{\hat{\rho}_\alpha(\vecr)}{\Phi_0}$. 
$\kappa_{\alpha\beta}(\vecr)$ is the residual interaction given as the second derivatives of the functional:
\begin{equation}
\kappa_{\alpha\beta}(\vecr)=\frac{\del^2 E_x [\left\{ \rho \right\}]}{\del \rho_\alpha(\vecr)\del\rho_\beta(\vecr)}.
\end{equation}

Note that the source of $\ket{\delta \Phi(t)}$ is not only the external field $\Vhat_{{\rm ext}}(t)$
but also the induced field $\delta\hat{V}_{{\rm ind}}(t)$, as indicated by Eq.(\ref{fluc-state}). 
Their sum is called the {\it selfconsistent
field}\cite{Zangwill-Soven,Nakatsukasa-Yabana}, 
which is given in the frequency domain 
as
\begin{equation}
\Vhat_{{\rm scf}} (\omega) \equiv \Vhat_{{\rm ext}}+\delta\Vhat_{{\rm ind}}(\omega) 
= \sum_\alpha  \int d\vecr 
v^{{\rm scf}}_\alpha(\vecr,\omega)\rhohat_\alpha(\vecr),
\end{equation}
\begin{equation}\label{SCF}
v_\alpha^{{\rm scf}}(\vecr,\omega) =
v_\alpha^{{\rm ext}}(\vecr) + 
\sum_\gamma\kappa_{\alpha\gamma}(\vecr)
          \delta\rho_\gamma(\vecr,\omega).
\end{equation}

The fluctuating densities $\delta\rho_\alpha(\vecr,t)$ are governed in the frequency domain
 by 
the linear response equation:
\begin{equation}\label{respeq1}
\delta\rho_\alpha(\vecr,\omega)=\int d\vecr' \sum_\beta
R_0^{\alpha\beta}(\vecr,\vecr',\omega)\left(
\sum_\gamma\kappa_{\beta\gamma}(\vecr')
          \delta\rho_\gamma(\vecr',\omega)
+v_\beta^{{\rm ext}}(\vecr')\right).
\end{equation}
Here $R_0^{\alpha\beta}(\vecr,\vecr',\omega)$ is the unperturbed response function,
which is expressed in the spectral representation as
\begin{equation}
R_0^{\alpha\beta}(\vecr,\vecr',\omega)=\sum_{i<j}
\left[ \frac{ \braket{0}{\hat{\rho}_\alpha(\vecr)}{ij} \braket{ij}{\hat{\rho}_\beta(\vecr')} {0} }
{\hbar\omega + i\epsilon - E_i - E_j}
- 
\frac{ \braket{0}{\hat{\rho}_\beta(\vecr')}{ij} \braket{ij}{\hat{\rho}_\alpha(\vecr)} {0} }
{\hbar\omega + i\epsilon + E_i + E_j}.
\right]
\end{equation}
Here use is made of  the quasiparticle states $i$ and $j$, which are Fermionic elementary
modes of the static HFB Hamiltonian $\hat{h}_0=\That+\Vhat_{{\rm mf}}[\{\rho_0\}]$, defined by
$[\hat{h}_0, a_i^\dagger] = E_i a_i^\dagger$. $\ket{0}$ denotes the HFB ground
state $\ket{\Phi_0}$, and $\ket{ij} \equiv a_i^\dagger a_j^\dagger \ket{\Phi_0}$ are two-queasiparticle
states.

The strength function
is given in terms of the 
 density response as
\begin{eqnarray}\label{strfn}
S(\hbar\omega)=-\frac{1}{\pi}{\rm Im}\int d\vecr \sum_\alpha 
\bar{v}_{\alpha}^{{\rm ext}}(\vecr)\delta\rho_{\alpha}(\vecr,\omega).
\end{eqnarray}
with $ \bar{v}_{\alpha}^{{\rm ext}}(\vecr)={v}_{\alpha}^{{\rm ext}}(\vecr)^* s_\alpha$.

\subsection{Partial cross sections for specific decay channels}

After absorbing a photon, the excited nucleus may decay by emitting
one or multiple nucleon(s) if the excitation energy is larger than the
threshold energies for the particle emissions. We shall formulate here a method
to evaluate 
partial cross sections of the photo-absorption reaction defined for specific
decay channels. To this end, we extend the method of
 Zangwill and Soven\cite{Zangwill-Soven} that is originally formulated for the continuum
RPA theory neglecting the pair correlations.  We shall show here
that the scheme can be generalized to the case of superfluid nuclei by
using the Bogoliubov quasiparticles instead of the single-particle states.

The starting point of the method is to note that 
the strength Eq.(\ref{strfn}) is rewritten
as
\begin{equation}
S(\hbar\omega)  =
-{1\over \pi}{\rm Im}\int\int d\vecr d\vecr' 
\sum_{\alpha\beta}
\bar{v}^{{\rm scf}}_{\alpha}(\vecr,\omega)
R_0^{\alpha\beta}(\vecr,\vecr',\omega)
v_{\beta}^{{\rm scf}}(\vecr',\omega)
 \label{strfn3} 
\end{equation}
in terms of the selfconsistent field and the unperturbed response
function. The derivation is given in Appendix A. Using the spectral
representation for $R_0(\omega)$, it is further written as
\begin{eqnarray}
S(\hbar\omega) 
&&=
-{1\over \pi}{\rm Im}\sum\sum_{i>j}
\left\{ 
{|\bra{ij}\Vhat_{{\rm scf}}(\omega)\ket{0}|^2 
\over \hbar\omega +i\eps-E_i-E_j }  
-
{|\bra{0}\Vhat_{{\rm scf}}(\omega)\ket{ij}|^2 
 \over \hbar\omega +i\eps+E_i+E_j}
\right\} \nonumber\\
&&= 
\sum\sum_{i>j} |\bra{ij}\Vhat_{{\rm scf}}(\omega)\ket{0}|^2 
\delta_\eps(\hbar\omega-E_{ij}) 
- |\bra{0}\Vhat_{{\rm scf}}(\omega)\ket{ij}|^2 
\delta_\eps(\hbar\omega+E_{ij})
\label{strfn4}
\end{eqnarray}
with a Lorentz function 
\begin{equation}
\delta_\eps(\hbar\omega\mp E_{ij})\equiv \frac{1}{\pi}
\frac{\eps}{(\hbar\omega  \mp E_{ij})^2 + \eps^2}, 
\ \ \ \ 
E_{ij}=E_i + E_j .
\end{equation}
We here recall (see Eq.(\ref{fluc-state}))  that the time-dependent field causing evolution of the system includes 
not only the external field $\hat{V}_{{\rm ext}}(t)=\Vhat_{{\rm ext}}e^{-i\omega t}+\Vhat_{{\rm ext}}^\dagger e^{i\omega t}$
but also the induced field $\delta\Vhat_{{\rm ind}}(t)$. This points to that 
$\bra{ij}\Vhat_{{\rm scf}}(\omega)\ket{0}$ is the matrix element for transition from the
ground state $\ket{0}$ to a two-quasiparticle state $\ket{ij}$. 
 If we take the limit $\eps \rightarrow 0$ in which 
$\delta_\eps(E)$ converges to the delta function $\delta(E)$, then we may interpret that 
each term of Eq.(\ref{strfn4}) is proportional to
\begin{equation}
w_{ij}=\frac{2\pi}{\hbar} 
 |\bra{{ij}}\Vhat_{{\rm scf}}(\omega)\ket{0}|^2 
\delta(\hbar\omega-E_{ij}), 
\end{equation}
which represents the transition probability per unit time  from the 
 the HFB ground state $\ket{0}$ to a two-quasiparticle state $\ket{ij}$. 
However, we need to pay attentions to 
spectral properties of the quasiparticle and two-quasiparticle states in order
to give precise interpretations to individual terms.

The quasiparticle
eigenstates of the HFB Hamiltonian $\hhat_0$ are categorized as either  discrete bound states or 
continuum unbound states\cite{DobHFB1,DobHFB2}. The discrete bound states are states satisfying
 $E_i < |\lambda|$ with $\lambda$ being the Fermi energy, and they  correspond 
to bound single-particle orbits which are located
 around the Fermi energy. We label them with $m,n$ etc. in the following.
 Those with $E_i > |\lambda|$ are all unbound states belonging to 
a continuum spectrum, and they describe a scattering nucleon. 
For the continuum quasiparticle states
we use labels $p(E_p),q(E_q)$ etc. with explicit quasiparticle energy.
Note that a part of single-particle hole orbits is embedded in the continuum spectrum due
to the coupling caused by the pair potential. Such hole-like quasiparticle states are
resonances in the HFB model.

Two-quasiparticle states $\ket{ij}$
are categorized in
three groups. The first is configurations, labeled $\ket{mn}$, in which two quasiparticles
are both discrete bound states. The second is configurations  $\ket{mp(E_p)}$ in which
one quasiparticle is in a bound state $m$ while the other is unbound continuum state $p(E_p)$.
They have a threshold energy
$S_1=|\lambda|+ \min E_m$,  i.e.,  the one-particle separation energy.
The third is the configurations $\ket{p(E_p)q(E_q)}$ with two particles in the continuum, 
and the corresponding threshold energy is the two-particle separation energy $S_2=2|\lambda|$.
Note that $S_1 \le S_2$.

We decompose the strength function according to these categories as
\begin{equation}\label{strfn-d1c2c}
S(\hbar\omega)=S_{d}(\hbar\omega)+S_{1c}(\hbar\omega)+S_{2c}(\hbar\omega)
\end{equation}
with
\begin{eqnarray}
S_{d}(\hbar\omega)=&& 
\sum\sum_{n>m} \left[ |\bra{nm}\Vhat_{{\rm scf}}(\omega)\ket{0}|^2 
\delta_\eps(\hbar\omega-E_{n}-E_{m}) \right.  \cr
&& \hspace{30mm}
\left. -|\bra{0}\Vhat_{{\rm scf}}(\omega)\ket{nm}|^2 
\delta_\eps(\hbar\omega+E_{n}+E_{m})\right] , \label{strfn-d}
 \\ 
S_{1c}(\hbar\omega)=&& 
\sum_{n}\SumInt_{p} \left[ |\bra{np(E_p)}\Vhat_{{\rm scf}}(\omega)\ket{0}|^2 
\delta_\eps(\hbar\omega-E_{n}-E_{p}) \right.  \cr
&& \hspace{30mm}
\left. - |\bra{0}\Vhat_{{\rm scf}}(\omega)\ket{np(E_p)}|^2 
\delta_\eps(\hbar\omega+E_{n}+E_{p})\right] , \label{strfn-1c}
 \\ 
S_{2c}(\hbar\omega)=&& 
\SumInt\SumInt_{p>q} \left[ |\bra{p(E_p)q(E_q)}\Vhat_{{\rm scf}}(\omega)\ket{0}|^2 
\delta_\eps(\hbar\omega-E_{p}-E_{q})\right.  \cr
&& \hspace{30mm}
\left. -  |\bra{0}\Vhat_{{\rm scf}}(\omega)\ket{p(E_p)q(E_q)}|^2 
\delta_\eps(\hbar\omega+E_{p}+E_{q})\right] . \label{strfn-2c}
\end{eqnarray}
Here $\SumInt_p=\sum'_p \int_{|\lambda|} dE_p$ denotes a summation over continuum quasiparticle states. 
Adopting
the partial wave representation, it is the integral over the quasiparticle energy $E_p > |\lambda|$ and
a summation $\sum'_p$ with respect to the angular quantum numbers. 
We shall  then examine properties of individual terms for
physical energies $E_\gamma=\hbar\omega >0$.

Let us first consider $S_{1c}(\hbar\omega)$ which collects contributions of two-quasiparticle
configurations $\left\{ \ket{np(E_p)} \right\}$. These configurations consist of 
one quasiparticle  in a scattering state $p(E_p)$ and the remaining odd-$A$ nucleus in a
one-quasiparticle state  $\ket{n}=a_n^\dagger \ket{0}$.  Now consider excitation energy 
$\hbar\omega > E_n+|\lambda|$ larger than the
threshold of this configuration. Then the integral range of $\int_{|\lambda|} dE_p$ includes the peak 
$E_p=\hbar\omega -E_n$ of
the Lorentz function $\delta_\eps(\hbar\omega -E_n -E_p)$ and hence  
$\delta_\eps(\hbar\omega -E_n -E_p)$ can be treated as the delta function 
while  $\delta_\eps(\hbar\omega +E_n +E_p)$ gives a contribution vanishing in the limit $\eps \rightarrow 0$.
Therefore  we find that each summand of $S_{1c}(\hbar\omega)$ in Eq.(\ref{strfn-1c})
 represents the probability of populating the one-quasiparticle state $\ket{n}$ and  one particle emitted
in the continuum scattering state $p(E_p)$. With multiplication of
 the kinematical factor $f(E_\gamma)$,  it is equal to the partial photo-absorption cross section 
\begin{equation}
\sigma_{\gamma \rightarrow np}(E_\gamma,E_p) =f(E_\gamma) |\bra{np(E_p)}\Vhat_{{\rm scf}}(\omega)\ket{0}|^2 
\delta(E_\gamma-E_n-E_p) 
\end{equation}
for  one-particle emission decay with the configurations mentioned above. 
Integrating
over the energy $E_p$, this term gives the on-shell cross section 
\begin{equation}
\sigma_{\gamma \rightarrow np}(E_\gamma)
=f(E_\gamma) |\bra{np(E_p)}\Vhat_{{\rm scf}}(\omega)\ket{0}|^2_{E_p=E_\gamma-E_n}.
\end{equation}
These are, in other words, the partial cross sections for
one-particle photo-dissociation.

Concerning $S_{2c}(\hbar\omega)$, each summand of Eq.(\ref{strfn-2c})
 represents a partial cross section for a
decay with emission of two particles in the scattering states $p(E_p)$ and $q(E_q)$:
\begin{equation}
\sigma_{\gamma \rightarrow pq}(E_\gamma,E_p,E_q) =f(E_\gamma)
|\bra{q(E_p)p(E_p)}\Vhat_{{\rm scf}}(\omega)\ket{0}|^2 
\delta(E_\gamma-E_p-E_q) .
\end{equation}

It should be noted that 
$S_{1c}(\hbar\omega)$ and $S_{2c}(\hbar\omega)$ have additional contribution in the energy region of
discrete spectrum $0< \hbar\omega <S_{1}=|\lambda| + \min E_n$ below the threshold $S_1$. This is because
the selfconsistent field $\Vhat_{{\rm scf}}(\omega)$
contains poles at the QRPA discrete eigen frequencies $\omega=\omega_k$ via the density response
$\delta\rho_\alpha(\vecr,\omega)$, and these pole contributions
give rise to the delta function peaks $\propto \delta(\hbar\omega-\hbar\omega_k)$.

The first term $S_d(\hbar\omega)$ has slightly different structure since the 
relevant two-quasiparticle configurations
$mn$ have discrete energies $E_n+E_m$. It might seem that this term exhibits discrete peaks
at energies $\hbar\omega=E_n+E_m$, but this is not
the case. As discussed in Appendix B, 
 $S_{d}(\hbar\omega)$   for $\hbar\omega > S_{1}$ (above the threshold energy) 
vanishes in the limit $\eps \rightarrow 0$. For $\hbar\omega < S_{1}$ (below the threshold), 
 $S_d(\hbar\omega)$ 
gives rise to discrete peaks $\propto \delta(\hbar\omega-\hbar\omega_k)$ as $S_{1c}(\hbar\omega)$
and  $S_{2c}(\hbar\omega)$ do.

Summarizing, Eqs. (\ref{strfn3}) and (\ref{strfn4}) of the strength function enables 
us to decompose the total photo-absorption
cross section, Eq.(\ref{totphotoabs}), into the partial photo-absorption cross sections 
associated with one- and two-particle
emission decays. Taking the limit $\eps \rightarrow 0$, we have
\begin{eqnarray}
\sigma_\gamma(E_\gamma) &=& f(E_\gamma)S(E_\gamma) \nonumber \\
&=&  \sum_{k,\hbar\omega_k <E_{th}}  \sigma_k \delta(E_\gamma-\hbar\omega_k)
+\sum_n\sum'_{p} \sigma_{\gamma \rightarrow np}(E_\gamma) 
+\SumInt\SumInt_{p>q} \sigma_{\gamma \rightarrow pq}(E_\gamma,E_p,E_q)  
\end{eqnarray}
where the partial cross sections  $\sigma_{\gamma \rightarrow np}$
and $\sigma_{\gamma \rightarrow pq}$
for the one- and two-particle emissions 
are, apart from the kinematical factor $f(E_\gamma)$, the summands of $S_{1c}(\hbar\omega)$ and 
  $S_{2c}(\hbar\omega)$ (Eqs.(\ref{strfn-1c}) and (\ref{strfn-2c})), respectively,
 while the photo-absorption cross
sections $\sigma_k$ to populate the bound excited states $k$ contain contributions from
all of $S_d(\hbar\omega), S_{1c}(\hbar\omega)$ and 
  $S_{2c}(\hbar\omega)$.

\subsection{Representation using wave functions and Green's function of quasiparticles }

It is useful to write down the above equations in terms of the quantities in 
the quasiparticle space and its coordinate representation.

A quasiparticle state has a two-component wave function\cite{DobHFB1,Matsuo01}
\begin{equation} \label{spinor}
\phi_i(\vecrs ) \equiv 
\left(
\begin{array}{c}
\vphi_{1,i}(\vecrs ) \\
\vphi_{2,i}(\vecrs )
\end{array}
\right).
\end{equation}
The matrix elements of the one-body operators $\hat{\rho}_\alpha(\vecr)$
are given as
\begin{eqnarray}
&\bra{ij}\rhohat_\alpha(\vecr)\ket{0}
  =\sum_\sigma \phi_i^\dag(\vecrs)\calA_\alpha \phicarj(\vecrs),  
\\
&\bra{0}\rhohat_\alpha(\vecr)\ket{ij}
  =\sum_\sigma \phicarj^\dag(\vecrs)\calA_\alpha \phi_i(\vecrs), 
\end{eqnarray}
in terms of $\phi_i(\vecrs )$ and its conjugate
\begin{equation} \label{conjugate}
\phicari(\vecrs ) \equiv 
\left(
\begin{array}{c}
-\vphi^*_{2,i}(\vecrs ) \\
\vphi^*_{1,i}(\vecrs)
\end{array}
\right) ,
\end{equation}
where $\phi_i(\vecrs )$ is the eigen wave function of the $2\times 2$ HFB Hamiltonian
$\calHz$ with the quasiparticle eigen energy $E_i$ while $\phicari(\vecrs )$
is the eigen function of the corresponding negative eigen energy $-E_i$. $\calA_\alpha$
is a local operator acting on the quasiparticle wave functions $\phi_i(\vecrs )$ and
 $\phicari(\vecrs)$. (Note that we follow the convention of the quasiparticle
wave function introduced in Ref.\cite{Matsuo01} except that
 $\phicari(\vecrs)$ in this paper replaces 
$ \phibari(\vecrs)$ used in the preceding papers\cite{Matsuo01,Serizawa09,Mizuyama09}.)

We shall use also the  quasiparticle Green's function
$
G(E)\equiv\left(E-\calHz\right)^{-1}
$
which is expressed in the spectral representation as
\begin{eqnarray}\label{green3}
G(\vecrs,\vecrsp,E)&&=
   \sum_n \left( {\phi_n(\vecrs)\phi_n^\dag(\vecrsp)\over E-E_n}
+  {\phicarn(\vecrs)\phicarn^\dag(\vecrsp)\over E+E_n} \right)  + G_c(\vecrs,\vecrsp,E),\nonumber\\
G_c(\vecrs,\vecrsp,E)&&=   \SumInt_p \left( {\phi_p(\vecrs)\phi_p^\dag(\vecrsp)\over E-E_p}
+  {\phicarp(\vecrs)\phicarp^\dag(\vecrsp)\over E+E_p} \right),
\end{eqnarray}
where $G_c(E)$ is a part arising from the continuum quasiparticle states.

The partial strength function $S_{1c}(\hbar\omega)$ associated with the one-particle continuum configurations $\ket{np(E_p)}$ is then written as  
\begin{eqnarray}
S_{1c}(\hbar\omega)=&&-{1 \over \pi}{\rm Im}\sum_n  \nonumber\\
&&\int\int d\vecr d\vecr'\sum_{\sigma\sigma'}
\left\{ \phicarn^\dagger(\vecrs)({\calV}_{{\rm scf}}(\vecr,\omega))^\dagger
{G}_{>}(\vecrs\vecrsp,\hbar\omega+i\eps-E_n)
\calV_{{\rm scf}}(\vecr',\omega)\phicarn(\vecrsp) \right. \nonumber \\
&& \hspace{10mm} +
\left. 
\phicarn^\dagger(\vecrsp)\calV_{{\rm scf}}(\vecr',\omega)
{G}_{>}(\vecrsp\vecrs,-\hbar\omega-i\eps-E_n)
({\calV}_{{\rm scf}}(\vecr,\omega))^\dagger\phicarn(\vecrs) \right\} \\ \label{strfn-1c3}
=&& 
-{1 \over \pi}{\rm Im}\sum_n 
   \left\{
\rbra{ \phicarn } (\calV_{{\rm scf}}(\omega))^\dagger 
 {G}_{>}(\hbar\omega+i\eps-E_n)
\calV_{{\rm scf}}(\omega) \rket{ \phicarn} \right.  \nonumber\\
&& \hspace{20mm} \left. +
\rbra{ \phicarn } \calV_{{\rm scf}}(\omega)
{G}_{>}(-\hbar\omega-i\eps-E_n)
({\calV}_{{\rm scf}}(\omega))^\dagger\rket{\phicarn} \right\}, \label{strfn-1c4}
\end{eqnarray}
where
\begin{equation}
\calV_{{\rm scf}}(\vecr,\omega)\equiv \sum_\alpha v_{\alpha}^{{\rm scf}}(\vecr,\omega)\calA_\alpha
\end{equation}
is the self-consistent field acting in the quasiparticle space.  We have introduced
a shorthand bra-ket notation in the last line of
 Eq.(\ref{strfn-1c4}). $G_>(E)$ is a part of $G_c(E)$ associated with 
the positive energy continuum states:
\begin{equation}
{G}_{>}(\vecrs,\vecrsp,E)\equiv 
\SumInt_p  {\phi_p(\vecrs)\phi_p^\dag(\vecrsp)\over E-E_p}.
\end{equation}

\begin{figure}[t]
\includegraphics[width=8cm,angle=0]{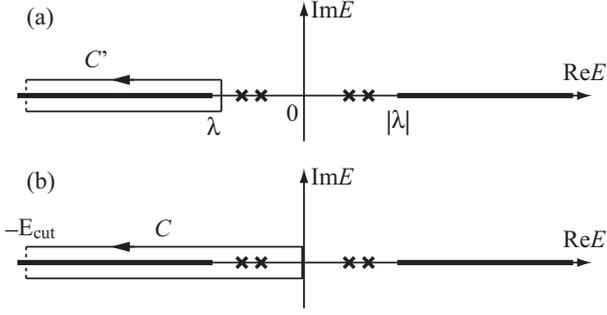}
\caption{The contours $C'$ and $C$ in the  complex quasiparticle energy space $E$, adopted for the integrations 
in Eqs.(\ref{str-2c-g}) 
and (\ref{contresp}). The crosses represent the poles at
$E=\pm E_i$ corresponding to the bound quasiparticle states. The
thick lines are the branch cuts corresponding to the continuum quasiparticle states.}\label{fig:contour}
\end{figure}

The strength function for the two-particle continuum is expressed as
\begin{eqnarray}\label{str-2c-g}
S_{2c}(\hbar\omega)  =-{1 \over \pi}{\rm Im}
{1\over 4\pi i}\int_{C'} dE 
&&\left\{ 
\Tr ({\calV}_{{\rm scf}}(\omega))^\dagger {G}_c(E+\hbar\omega+i\eps)
        \calV_{{\rm scf}}(\omega) G(E)  \right. \nonumber\\
&&\left. +\Tr ({\calV}_{{\rm scf}}(\omega))^\dagger G(E)
       \calV_{{\rm scf}}(\omega) {G}_c(E-\hbar\omega-i\eps) \right\}.
\end{eqnarray}
using the complex energy integration along the contour $C'$ shown in Fig.\ref{fig:contour}(a).

\subsection{Partial cross sections for one-particle decays}

Let us concentrate on the  partial cross section for one-particle decay channels to give
a concrete expression to be used in numerical calculation.

We rewrite Eq.(\ref{strfn-1c4}) as
\begin{equation}\label{strfn-1cp1}
S_{1c}(\hbar\omega)=
-{1 \over \pi}{\rm Im}\sum_n
 \rbra{\phicarn}(\calV_{{\rm scf}}(\omega))^\dagger
{G}_c(\hbar\omega-E_n+i\eps)
\calV_{{\rm scf}}(\omega)\rket{\phicarn}  + \Delta S_{1c}(\hbar\omega)
\end{equation}
with 
\begin{eqnarray}\label{strfn-1cp2}
\Delta S_{1c}(\hbar\omega) 
&=&-\sum_n\SumInt_p \left|\rbra{\phicarn}\calV_{{\rm scf}}(\omega)\rket{\phi_p}\right|^2 \delta_\eps(\hbar\omega - E_n +E_p)
\nonumber\\
&&-\sum_n\SumInt_p \left|\rbra{\phicarp}\calV_{{\rm scf}}(\omega)\rket{\phicarn}\right|^2 \delta_\eps(\hbar\omega + E_n +E_p).
\end{eqnarray}
We remark that 
the second term $\Delta S_{1c}(\hbar\omega)$  vanishes if we take the limit $\eps \rightarrow 0$
and as far as we consider the excitation energies $\hbar\omega > S_{1}$ above the
one-particle separation energy $S_1=\min E_n + |\lambda|$. This is because
$\hbar\omega \mp E_n+ E_p >2|\lambda| - \max E_n >|\lambda|$, and hence
$\delta_\eps(\hbar\omega \mp E_n + E_p) \propto \eps \rightarrow 0$.

In the following we assume that the mean fields in the  HFB Hamiltonian $\calHz$
is spherically symmetric. We use the partial wave expansion:
$\phi_{nljm}(\vecrs)=r^{-1}\phi_{nlj}(r)\calY_{ljm}(\hat{\vecr}\sigma)$  for the bound quasiparticle states,
 and
$G(\vecrs,\vecrsp,E)=\sum_{l'j'm'}(rr')^{-1}G_{l'j'}(r,r',E)\calY_{l'j'm'}^\dagger (\hat{\vecr}\sigma)\calY_{l'j'm'}^\dagger (\hat{\vecr}'\sigma')$ for the quasiparticle Green's function, where
$ljm$ and $n$ are angular  and radial  quantum  numbers, respectively.
Using these quantum numbers, the partial cross section $\sigma_{\gamma\rightarrow np}(E_\gamma)$ for one-particle
decay is specified by the quantum number
$el'j'$ of the emitted nucleon in the continuum state (with kinetic energy $e=E-|\lambda|$)
and the quantum number $nlj$ of a bound one-quasiparticle
state of the remaining odd-$A$ nucleus, as well as the multipolarity $L$ of the gamma-ray.
The one-particle photo-dissociation cross section 
for this specific channel is given as
\begin{eqnarray}\label{strfn-1c-pw}
\sigma_{\gamma \rightarrow nlj, l'j'}^{L}(E_\gamma)
&=& -{ f(E_\gamma) \over \pi}  {\left<l'j'\right\|Y_L\left\|lj\right>^2 \over 2L+1} \nonumber\\
 && \hspace{-10mm} \times  {\rm Im} \int_0^{R_2} dr\int_0^{R_2} dr' 
\phibar_{nlj}^{T}(r)\left(\calV_{{\rm scf},L}(r,\omega)\right)^\dagger {G}_{c,l'j'}(r,r',\hbar\omega-E_{nlj}+i\eps)\calV_{{\rm scf},L}(r',\omega) 
\phibar_{nlj}(r')\nonumber\\
&& + \Delta\sigma_{\gamma \rightarrow nlj,l'j'}^{L}(E_\gamma)
\end{eqnarray}
where  $\calV_{{\rm scf},L}(r,\omega)$ is the radial component of the self-consistent field defined by
$
\calV_{{\rm scf}}(\vecr,\omega) = \calV_{{\rm scf},L}(r,\omega)Y_{LM}(\hat{\vecr}), 
$
and the continuum part of the Green's function can be calculated as
\begin{equation}
{G}_{c,lj}(r,r',E)= G_{lj}(r,r',E) - \sum_n 
\left\{ 
{ \phi_{nlj}(r)\phi_{nlj}^T(r') \over E-E_{nlj}} +  { \phibar_{nlj}(r)\phibar_{nlj}^T(r') \over E+E_{nlj}}
\right\}.
\end{equation}
by subtracting the contribution of the discrete quasiparticle states from
the exact HFB Green's function $G_{lj}(E)$. The Green's function  $G_{lj}(E)$ can be constructed
exactly in terms of quasiparticle wave functions regular  at the origin,
and quasiparticle wave function satisfying the boundary condition at infinity\cite{Belyaev87,Matsuo01}.
In practice we connect the latter to the Hankel functions at a large radius $R_2$ (see below).
[The Coulomb function should be used in the case of proton emission.]
 Here  $\Delta\sigma_{\gamma\rightarrow nlj,l'j'}^{L}(E_\gamma)$
corresponds to a summand in $\Delta S_{1c}$ in Eq.(\ref{strfn-1cp2}). We calculate this 
 unimportant term (vanishing in the limit $\eps \rightarrow 0$)
by replacing the exact continuum states with discretized continuum states obtained with
the box boundary condition at $r=R_2$.

\begin{figure}[t]
\includegraphics[width=8cm]{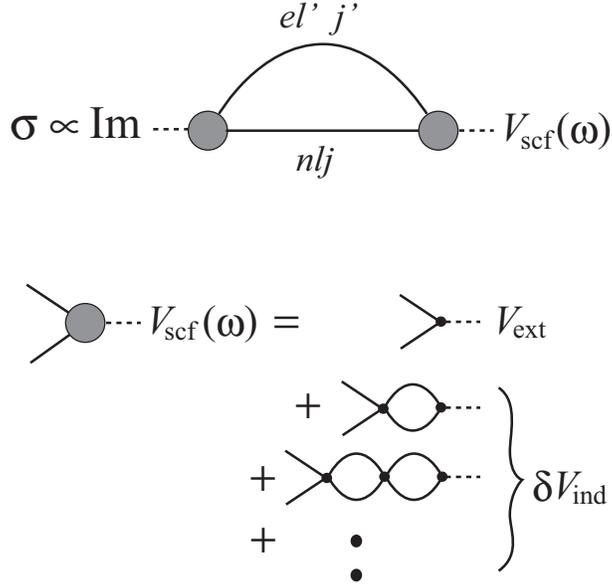}
\caption{A diagrammatic representation of the partial photo-absorption
cross section $\sigma_{\gamma \rightarrow nlj, l'j'}^{L}(E_\gamma)$,
Eq.(\ref{strfn-1c-pw}) and the vertex associated with the selfconsistent
field $V_{{\rm scf}}(\omega)$.
}
\label{fig:vertex}
\end{figure}

Equation (\ref{strfn-1c-pw}) is given a diagrammatic representation as in Fig.\ref{fig:vertex}. Note that the 
vertex to the photon is not a bare dipole operator $\Vhat_{{\rm ext}}=D_0$ but
the selfconsistent field $\Vhat_{{\rm scf}}(\omega)$ which includes 
the polarization effect caused by the correlation in nuclei via the induced field.

\subsection{Direct neutron capture cross section}

The inverse process of the photo-absorption reaction leading to a specific decay channel
$(nlj)(el'j')$ is the capture of a nucleon with kinetic energy $e$ in the partial wave $l'j'$ by 
the odd-$A$ nucleus with the one-quasiparticle configuration $(nlj)$, followed by the
photon emission populating the ground state of the fused even-$A$  nucleus. Using the 
reciprocity theorem, one can calculate the cross section of the inverse process with
\begin{equation}\label{ncapture}
\sigma^{{\rm cap},L}_{nlj+l'j' \rightarrow \gamma }(e)= \frac{1}{2j+1}\frac{E_\gamma^2}{2m c^2 e}\sigma_{\gamma \rightarrow nlj,l'j'}^L(E_\gamma),
\end{equation} 
where $e=E_\gamma - E_{nlj}$ is the nucleon kinetic energy and $m$ is the nucleon  mass. 
This is nothing but the
radiative capture cross section for a nucleon in the partial wave $l'j'$ captured 
by the odd-$A$ nucleus with the one-quasiparticle configuration $nlj$, followed by
an E1 transition to the $0^+$ ground state of the even-$A$ nucleus.

\section{Numerical procedure}

In the following we shall demonstrate the present theory with a numerical example.
For this purpose, we take neutron-rich tin isotopes with mass number $A\sim 140$,
in which one-neutron separation energy is $S_{1n} \sim $ a few MeV, and hence
the direct neutron capture is expected to be dominant in the r-process reaction 
of these isotopes\cite{Xu-Goriely14,Mathews83,Rauscher10}.
Another reason of the choice is that we can assume the spherical shape
because of the proton magicity $Z=50$ as many
Hartree-Fock-Bogoliubov calculations predict\cite{Stoitsov03,Dobaczewski04,Lala98}.

\begin{figure}[t]
\includegraphics[width=8cm,angle=-90]{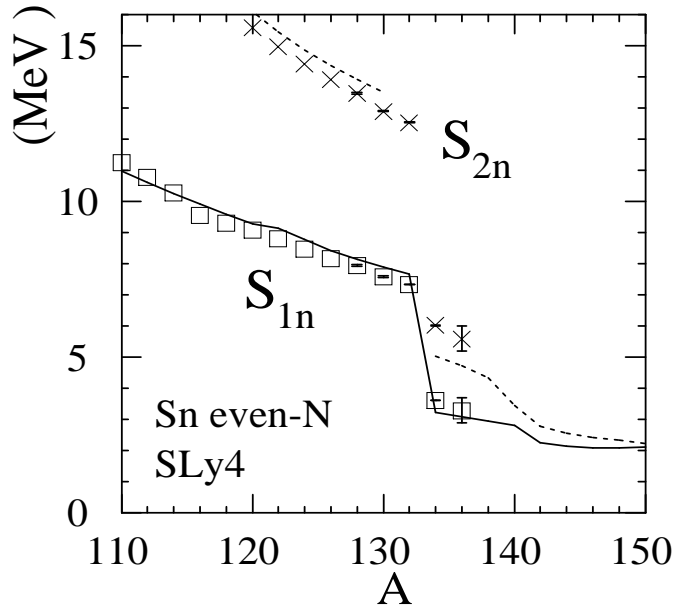}
\caption{
Calculated one- and two-neutron separation energies $S_{1n}$ and $S_{2n}$ for even-even 
Sn isotopes, plotted with solid and dotted lines, respectively. The experimental values\cite{NNDC}
are shown with squares and crosses for  $S_{1n}$ and $S_{2n}$, respectively.
}\label{fig:separation}
\end{figure}

We employ the Skyrme energy density functional model and the effective pairing interaction
of the contact type to construct the HFB ground state and the associated selfconsistent
mean-field. The adopted Skyrme parameter set is SLy4\cite{CBH98}, and the density-dependent
delta interaction of the mixed type\cite{DD-Dob,DD-mix1,DD-mix2}
\begin{equation}
v(1,2)=V_0\left(1-\frac{\rho(\vecr)}{2\rho_0}\right)\delta(\vecr_1-\vecr_2)
\end{equation}
($\rho_0=0.16$ fm$^{-3}$) is adopted with the quasiparticle energy cut-off $E_{{\rm cut}}=60$ MeV 
and the orbital
angular momentum cut-off $l_{{\rm cut}}=12$. The adopted force strength $V_0=292$ MeV fm$^{-3}$ 
gives   average neutron pairing gap $\Delta_n=1.25$ MeV 
 for stable isotope $^{120}$Sn and  $\Delta_n=0.97$ MeV for   $^{142}$Sn.
 The proton pairing gap is zero.
 The radial HFB equation is solved in a spherical box $r<R_1$ with mesh interval $\Delta r=0.2$ fm
under the box boundary
condition $\phi_i|_{r=R_1}=0$ with the box radius $R_1=20$ fm, which is sufficiently
large for bulk quantities such as the total binding energy to converge. The calculation
reproduces rather well the experimental 
one- and two-neutron separation energies, $S_{1n}$ and $S_{2n}$, of the even-$N$ tin isotopes,
 as shown in Fig.\ref{fig:separation}, and it
predicts $S_{1n} \approx 2 $MeV  for $A> 140$.

The continuum QRPA calculation is performed as follows.
We adopt the Landau-Migdal approximation in evaluating the linear response: we consider only the
fluctuations in the local density and local pair density, and we employ the
Landau-Migdal parameters $F_0(\vecr)$ and $G_0(\vecr)$ in the local density approximation for the 
particle-hole residual
interaction. 
We solve the linear response equation 
\begin{equation} \label{resp-rsp}
\delta\rho_{\alpha L}(r,\omega)=\int_0^{R_2}dr' \sum_\beta R_{0,L}^{\alpha\beta}(r,r',\omega)
\left( \sum_\gamma \kappa_{\beta\gamma}(r')\frac{1}{r^2}\delta\rho_{\gamma L}(r',\omega) + v^{{\rm ext}}_{\beta L}(r')
\right) 
\end{equation}
in the radial coordinate space.
For the unperturbed response function, 
we use the representation\cite{Matsuo01} using the quasiparticle Green's function:
\begin{eqnarray}\label{contresp}
 R_{0,L}^{\alpha\beta}(r,r',\omega)&&=\frac{1}{4\pi i} \int_C dE
\sum_{lj,l'j'} {\left<l'j'\right\|Y_L\left\|lj\right>^2 \over 2L+1}
\left\{ {\rm Tr}\calA_\alpha G_{l'j'}(r,r',E+\hbar\omega+i\eps)A_\beta G_{lj}(r',r,E) \right. \nonumber \\
&& \hspace{20mm} + \left. {\rm Tr}\calA_\alpha G_{lj}(r,r',E)A_\beta G_{l'j'}(r',r,E-\hbar\omega-i\eps)
\right\}
\end{eqnarray}
in order to treat the continuum quasiparticle states with the proper boundary condition.
The contour $C$ is the one shown in Fig.\ref{fig:contour}(b).

In finding a numerical solution of the linear response equation 
(\ref{resp-rsp}) (using a matrix form with radial mesh points),
and also performing numerical integration in Eq.(\ref{strfn-1c-pw}), we need 
a large radial space so that we can evaluate the matrix element
$\bra{np(E_p)}\Vhat_{{\rm scf}}(\omega)\ket{0}$ accurately for scattering state $p(E_p)$ and weakly
bound quasiparticle state $n$. We specify This space is specified with a radius  $R_2$.
We found that a choice $R_2=R_1=20$ fm  provides
reasonable results, which however are not sufficiently converged for the 
photo-absorption cross sections with
low-energy neutron emission. 
However, we cannot simply
enlarge the HFB cut-off radius $R_1$ since
numerical solution for a quasiparticle state with large quasiparticle energy exhibits an exponential growth
of error 
when the radial HFB equation is integrated toward large $r$ ($\gesim 25$ fm in the present case), as
discussed by Bennaceur and Dobaczewski\cite{BennaceurCPC}.
 To avoid this problem,
we adopt the following two prescriptions. 
i) For the cut-off radius $R_2$ used in Eqs. (\ref{resp-rsp}) and (\ref{strfn-1c-pw}), 
we choose a value larger than $R_1$,  while keeping
$R_1$ for the HFB calculation. We neglect  
the HFB mean-fields in calculating
wave functions in the interval $R_1 < r <R_2$  since this
potential cut-off is known to stabilize significantly the numerical solution of quasiparticle wave functions\cite{BennaceurCPC}.
The potential cut-off can be justified except for nuclei with very small binding energies $\ll 1$ MeV
and with very long
tails of the density and the pair density extended to far distances.
ii) We introduce a smaller cut-off energy $E_{{\rm cut,out}}$ for the upper boundary of the Contour integral in Eq.(\ref{contresp}) 
in evaluating
the unperturbed response function $R_{0,L}^{\alpha\beta}(r,r',\omega)$ at large distances so that the
above numerical problem does not come into play. In practice, we use $E_{{\rm cut,out}}=10$ MeV for
$r'>R_1$ while the original cut-off value $E_{{\rm cut}}=60$ MeV is used for $r'<R_1$. We found that this choice
gives a convergence with respect to $E_{{\rm cut,out}}$. 
We also
found that the convergence of the cross section with respect to $R_2$ is obtained with
$R_2=30$ fm.  In evaluating the unperturbed response function, necessary for the
QRPA calculation, we use a small but finite value of the imaginary constant
$\eps=0.05$ MeV, corresponding to a smearing width of $100$ keV in the strength
function.   

One needs to evaluate the partial photo-absorption cross sections and
the neutron capture cross sections with very fine energy resolution if one
wants to apply to the astrophysical problems since the relevant energy scale
of the neutron kinetic energy is $e \sim 1\times 10^{-3} - 1 $ MeV.
For this purpose,  we use a  very small
imaginary constant 
$\eps=1\times 10^{-8}$ MeV for the Green's function appearing in Eq.(\ref{strfn-1c-pw}), thus
allowing description of the neutron scattering states with  energy resolution
$\sim \eps$.

\section{Numerical example}

\begin{figure}[t]
\includegraphics[width=8cm,angle=-90]{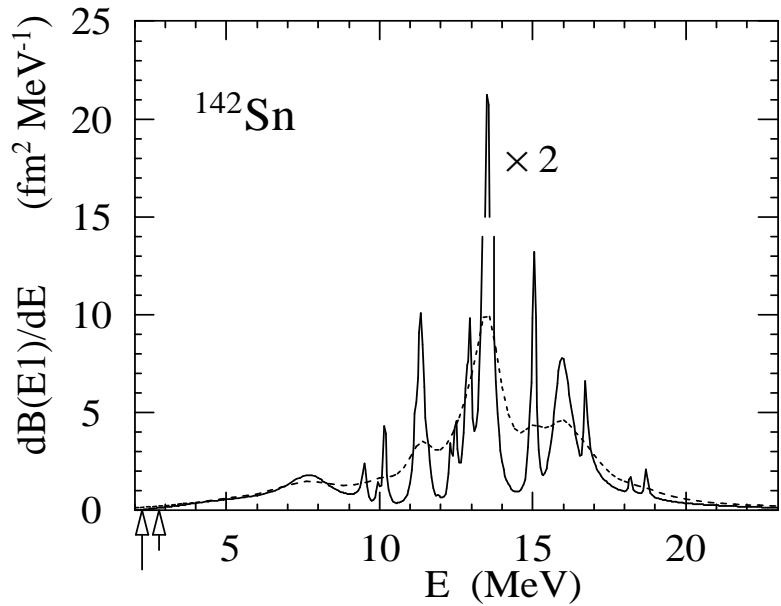}
\caption{
Calculated E1 strength function $dB(E1)/dE$ in $^{142}$Sn. The solid curve is the strength obtained with a smearing
width $\gamma=2\eps=100$ keV, while the dashed curve is for a smearing width of 1 MeV. 
The long and short arrows indicate the one- and two-neutron separation energies
$S_{1n}$ and $S_{2n}$, respectively.
}\label{fig:E1}
\end{figure}

Figure \ref{fig:E1} shows the calculated E1 strength 
$dB(E1)/dE\equiv 3\sum_k |\bra{k}D_0 \ket{0}|^2 
\delta(E-\hbar\omega_k) =3S(E)$ plotted as a function
of the excitation energy $E$. A large fraction of the strength
is distributed around $E \sim 10-17$ MeV,
corresponding to the giant dipole resonance. The
strength is also seen between $E \sim 10$ MeV and the one-neutron separation
energy $S_{1n}$, and it is of our interest in this study.
Significant fluctuation or fine structure in the GDR region is seen. They
reflect bound proton particle-hole configurations and neutron configurations
involving quasiparticle resonances with narrow width. 
These fine structures might disappear if we take into account  
the spreading width arising from 
coupling to more complex configurations, e.g. four quasiparticle configurations
etc.  If we simulate the spreading width using a finite value of the
smearing width,  the E1 strength strength distribution becomes smooth as
illustrated by the
dashed curve, obtained with the smearing width of $\gamma=1$ MeV.

\begin{figure}[t]
\includegraphics[width=8cm,angle=-90]{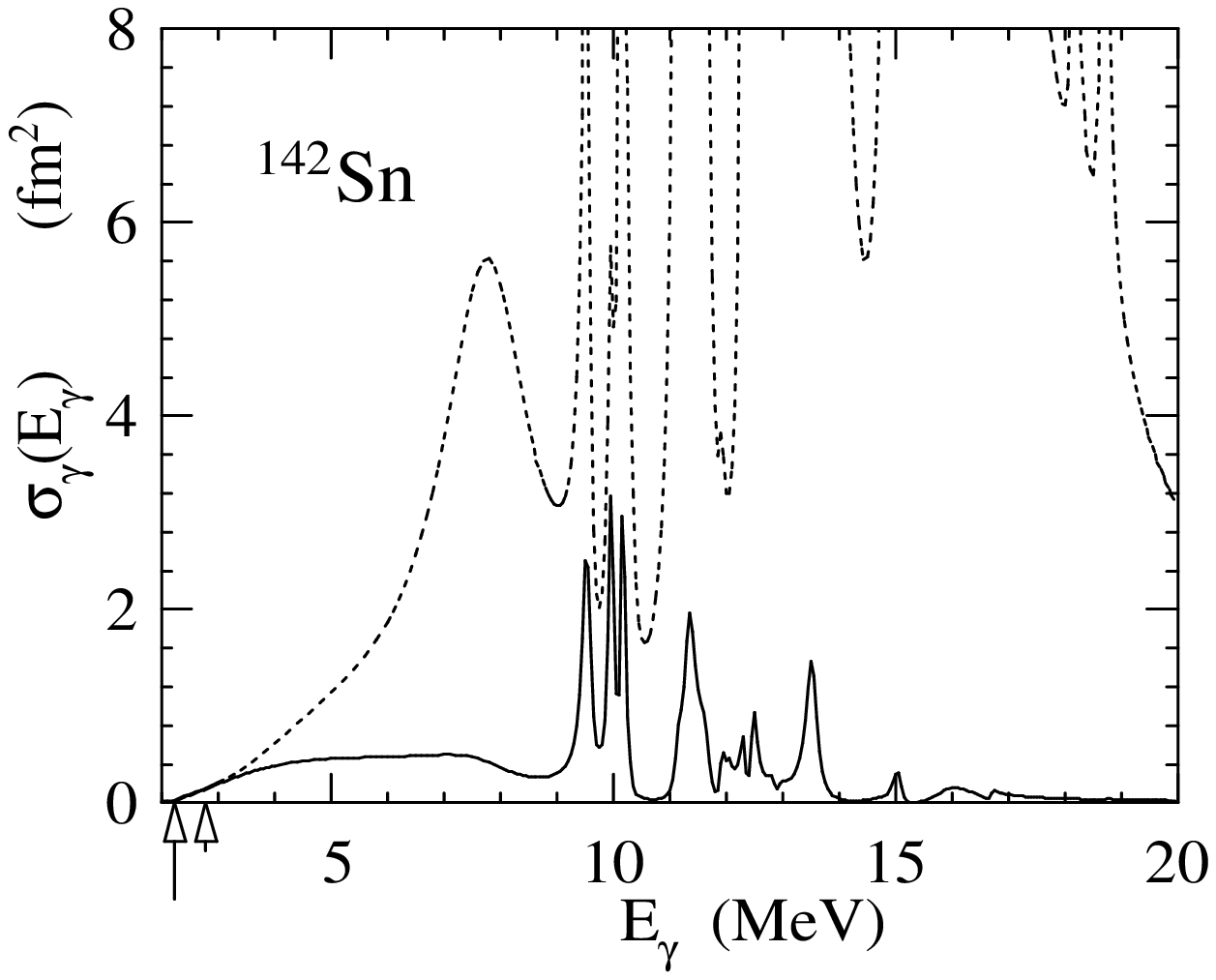}
\caption{
Total photo-absorption cross section and partial cross section for one-neutron emission decay,
plotted with dotted and solid curves, respectively,
calculated  for $^{142}$Sn.
} \label{fig:photo-abs-1}
\end{figure}

\begin{figure}[t]
\includegraphics[width=8cm,angle=-90]{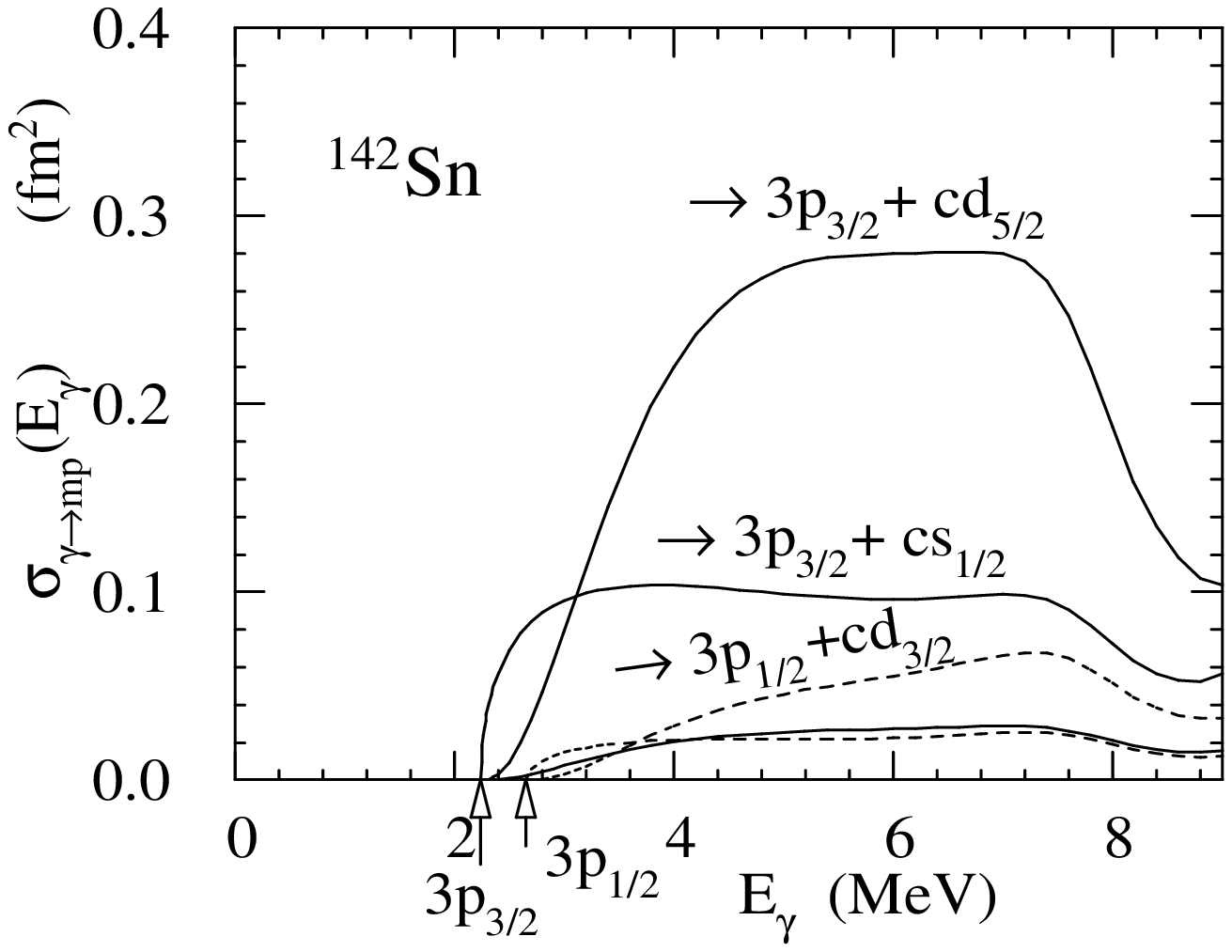}
\caption{
Partial photo-absorption cross sections for specific channels of one-neutron emission decay,
calculated for $^{142}$Sn. The three solid curves are for  channels
 $3p_{3/2}\otimes cs_{1/2}$,  $3p_{3/2}\otimes cd_{3/2}$ and  $3p_{3/2}\otimes cd_{5/2}$
while the two dashed curves are for  $3p_{1/2}\otimes cs_{1/2}$ and $3p_{1/2}\otimes cd_{3/2}$
(see text for details). The arrows indicate the threshold energies of these decay channels.
}\label{fig:photo-abs-2}
\end{figure}

Figures \ref{fig:photo-abs-1} and \ref{fig:photo-abs-2} show calculated total and partial photo-absorption
cross sections.  The total photo-absorption cross section $\sigma_\gamma(E_\gamma)$, the dotted
curve in Fig.\ref{fig:photo-abs-1}, is proportional to $E dB(E1)/dE |_{E=E_\gamma}$ and hence it
has basically the same structure as the E1 strength function. 
Open decay modes of the excited $1^-$ states
 are one- and two-neutron emissions. The one-
two-neutron separation energies are low: $S_{1n}= E_{3p_{3/2}} + |\lambda_n|=2.246$ MeV and
$S_{2n}= 2 |\lambda_n|=2.796$ MeV, respectively ($E_{3p_{3/2}}$ is the quasiparticle energy
of the neutron $3p_{3/2}$ state).  The one-proton separation energy
$S_{1p}=|e_{1g_{9/2}}|=18.191$ MeV is located at much higher energy. The partial
photo-absorption cross section 
 for one-neutron emission decay is shown with the solid curve in
Fig.\ref{fig:photo-abs-1}. It is seen that the partial cross section for two-neutron
 decay becomes a sizable fraction for  $E_\gamma \gesim 3.5$ MeV. The
fraction of one-neutron  cross section becomes significantly small as the energy increases
although the one-neutron decay survives at energies where the two-neutron  decay channels
are open. 
 We remark also that the one-neutron partial
cross section exhibits non-trivial energy dependence which arises from the configuration 
mixing in the dipole states. For instance, the fine structures seen around 
$E_\gamma \sim  9-17$ MeV can be explained only with mixing among
 proton particle-hole  and neutron two-quasiparticle configurations. 

\begin{figure}[t]
\includegraphics[width=8cm,angle=-90]{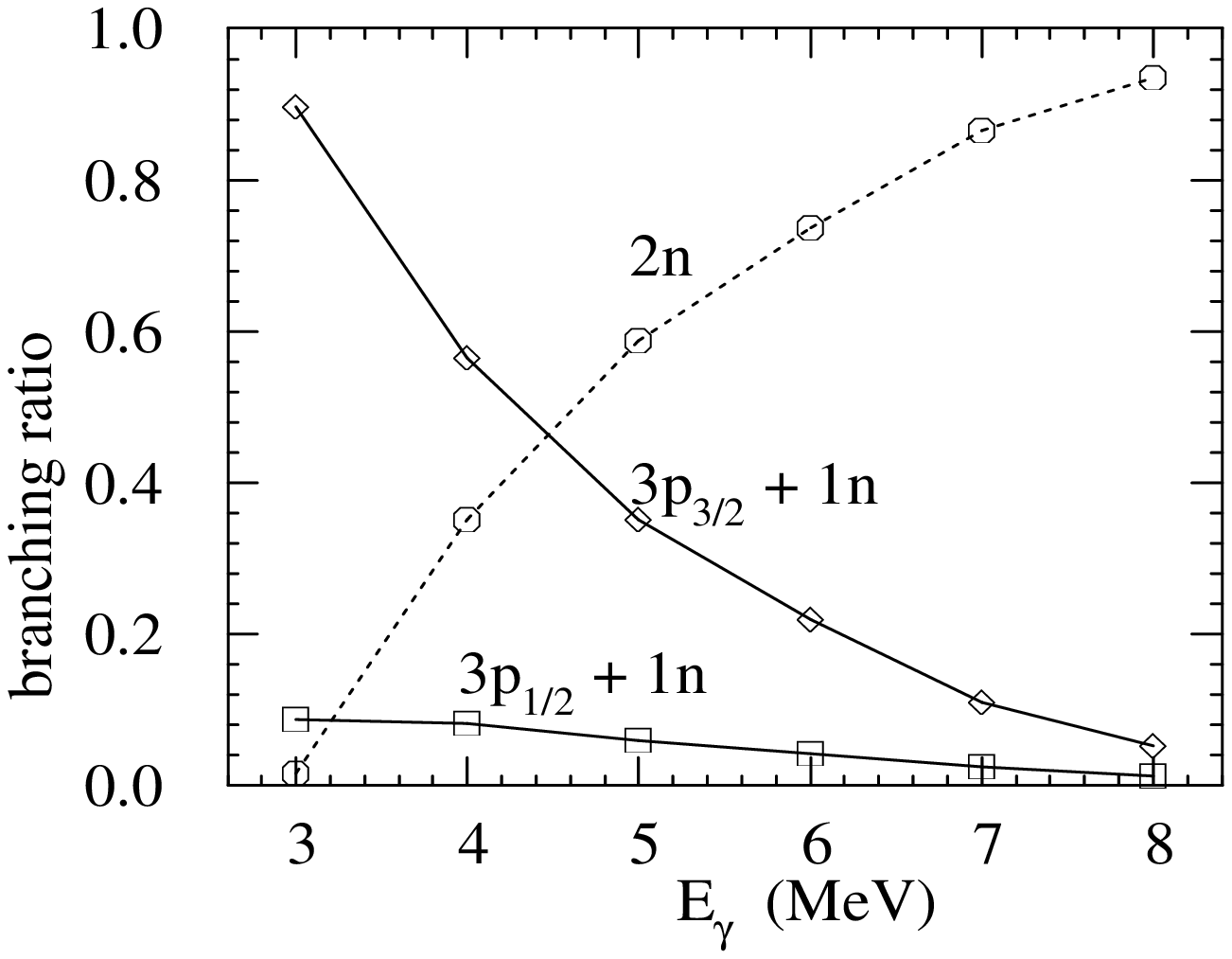}
\caption{
Calculated branching ratios for  one-neutron decays from photo-excited $1^{-}$ states
of $^{142}$Sn
populating the $3p_{3/2}$ state of the daughter $^{141}$Sn (plotted with diamonds),  
the same but for the  $3p_{3/2}$ state (squares), and that for two-neutron decays (circles),
evaluated for various photon energies $E_\gamma$.
}\label{fig:branching}
\end{figure}

The one-neutron decay is further decomposed into individual decay channels specified 
with different neutron configurations.   
In the present case bound neutron quasiparticle states are  
$3p_{3/2}$ and $3p_{1/2}$ states with quasiparticle energies 
$E_{3p_{3/2}}=0.848$ MeV and $E_{3p_{1/2}}=1.257$ MeV while all the other quasiparticle
states are in the continuum $E>|\lambda_n|$. Therefore configurations corresponding
to the final states of one-neutron decay are the $3p_{3/2}$ state coupled with
continuum $s_{1/2}$, $d_{5/2}$ and $d_{3/2}$ states, combined in total spin
 and parity $1^-$ (abbreviated as $3p_{3/2}\otimes cs_{1/2}$, $3p_{3/2}\otimes cd_{3/2}$
and  $3p_{3/2}\otimes cd_{5/2}$, hereafter), and  similarly,
$3p_{1/2}\otimes cs_{1/2}$ and  $3p_{1/2}\otimes cd_{3/2}$, involving the $3p_{1/2}$ state.
The first three are
decay channels in which the one-quasiparticle state  $3p_{3/2}$, 
 the calculated ground state of $^{141}$Sn, is populated while the last two are those
populating the 
 one-quasiparticle state  $3p_{1/2}$, the only bound excited state in  $^{141}$Sn
 obtained in the present HFB calculation.
The partial photo-absorption cross sections for these decay channels are
plotted in Fig.\ref{fig:photo-abs-2}. The decay channels with population of the ground 
$3p_{3/2}$ state
open at $E_\gamma=E_{3p_{3/2}}+|\lambda_n|=S_{1n}$ while the channels populating
the $3p_{1/2}$ state open at $E_\gamma=E_{3p_{1/2}}+|\lambda_n|=2.655$ MeV, higher
than $S_{1n}$ by 409 keV. It is seen that the probability of populating the 
excited  $3p_{1/2}$ state is finite but much smaller than that populating the ground state
 $3p_{3/2}$. We show in Fig.\ref{fig:branching} the decay 
branching ratio. It is seen that the branching ratio varies with excitation energy, displaying
 monotonic increase (decrease) of the two-neutron (one-neutron) decay branches.

Focusing on the ground state decays (the solid curves in Fig.\ref{fig:photo-abs-2}),
we find an apparent feature that the channel with the escaping neutron in the $s_{1/2}$ wave dominates
over those in the $d$ waves at the lowest energies close to
the threshold. At higher energies $E_\gamma \gesim 3$ MeV, the channel with the
$d$-wave neutron dominates. 

\begin{figure}[t]
\includegraphics[angle=-90,width=8cm]{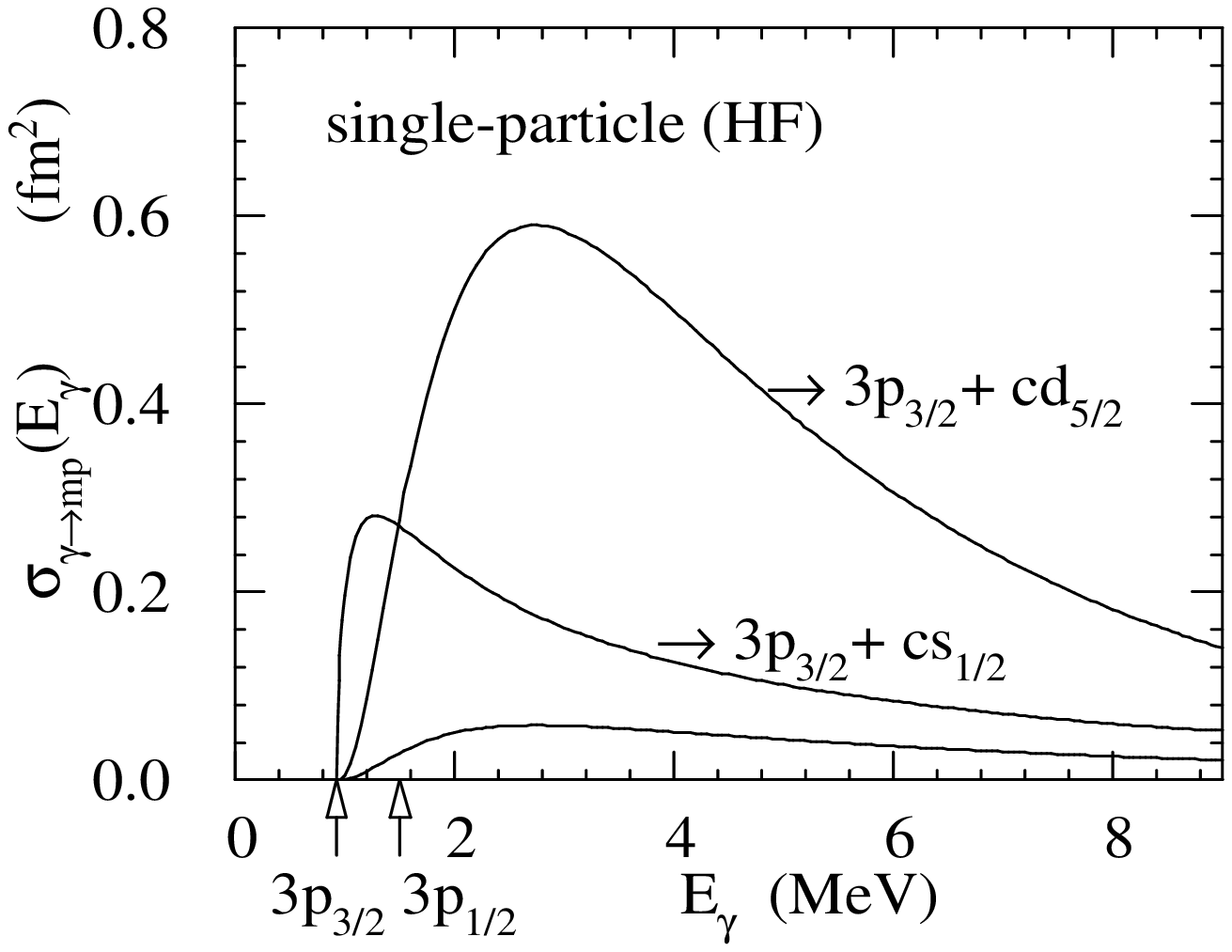}
\caption{
Partial photo-absorption cross sections 
 for specific channels of one-neutron emission decay, obtained
by neglecting 
the pairing and RPA correlations. Decays populating
the $3p_{3/2}$ and $3p_{1/2}$
states of the daughter $^{141}$Sn are evaluated, but  the cross sections
for the  $3p_{1/2}$ state is calculated to be zero in this null pairing case.
}\label{fig:photo-abs-sp}
\end{figure}

It is interesting to compare this result with a simple model corresponding to
single-particle transitions in the Hartree-Fock approximation. For the latter, we perform a calculation neglecting
 the pairing correlation and the RPA correlation caused by the residual interactions.
In practice we perform the HFB calculation using
 a reduced paring interaction strength $V_0=120$ MeV fm$^{-3}$, which leads to
a very small average neutron pairing gap $\Delta_n=0.048$ MeV, and calculate the
partial photo-absorption cross sections using Eq.(\ref{strfn-1c-pw}) in which the selfconsistent field
$V_{{\rm scf}}(\omega)$ is replaced with the bare dipole operator. The result is shown in 
Fig.\ref{fig:photo-abs-sp} for the decay channels populating the $3p_{3/2}$ and $3p_{1/2}$
state of the daughter $^{141}$Sn. (The cross section for the $3p_{1/2}$ state
is calculated to be practically zero.)

Several clear differences are seen between Figs.\ref{fig:photo-abs-2}
and \ref{fig:photo-abs-sp}. First, the one-neutron separation energy is higher
in the full calculation by about 1.5 MeV than that in the Hartree-Fock 
single-particle model.  This due to the pair correlation which has an effect to
give the even-$N$ nucleus $^{142}$Sn more binding energy. The separation energy in the
Hartree-Fock approximation is essentially the single-particle
energy -0.883 MeV of the $3p_{3/2}$ orbit while the pair correlation
increases the separation energy via
the quasiparticle energy $E_{3p_{3/2}}$ and the Fermi energy $\lambda_n$. 
Second, the probability
to populate the  $3p_{1/2}$ state is finite in the full HFB + QRPA calculation while it is zero in
the Hartree-Fock approximation. This is because the single-particle $3p_{1/2}$ orbit
is partially occupied in the HFB description of the pair correlated ground state of $^{142}$Sn
while the occupation is zero in the unpaired Hartree-Fock approximation.
Third, the cross sections have non-trivial energy dependence in the full calculation
while the energy-dependence in the Hartree-Fock single-particle transitions are
quite simple. The non-trivial energy dependence is due to the RPA correlation and
the configuration mixing as we already mentioned above. The simple structure 
in the single-particle transitions, on the other hand, can be understood even 
in an analytical way\cite{Typel05,Nagarajan05}.

\begin{figure}[t]
\includegraphics[clip,width=8cm,angle=-90]{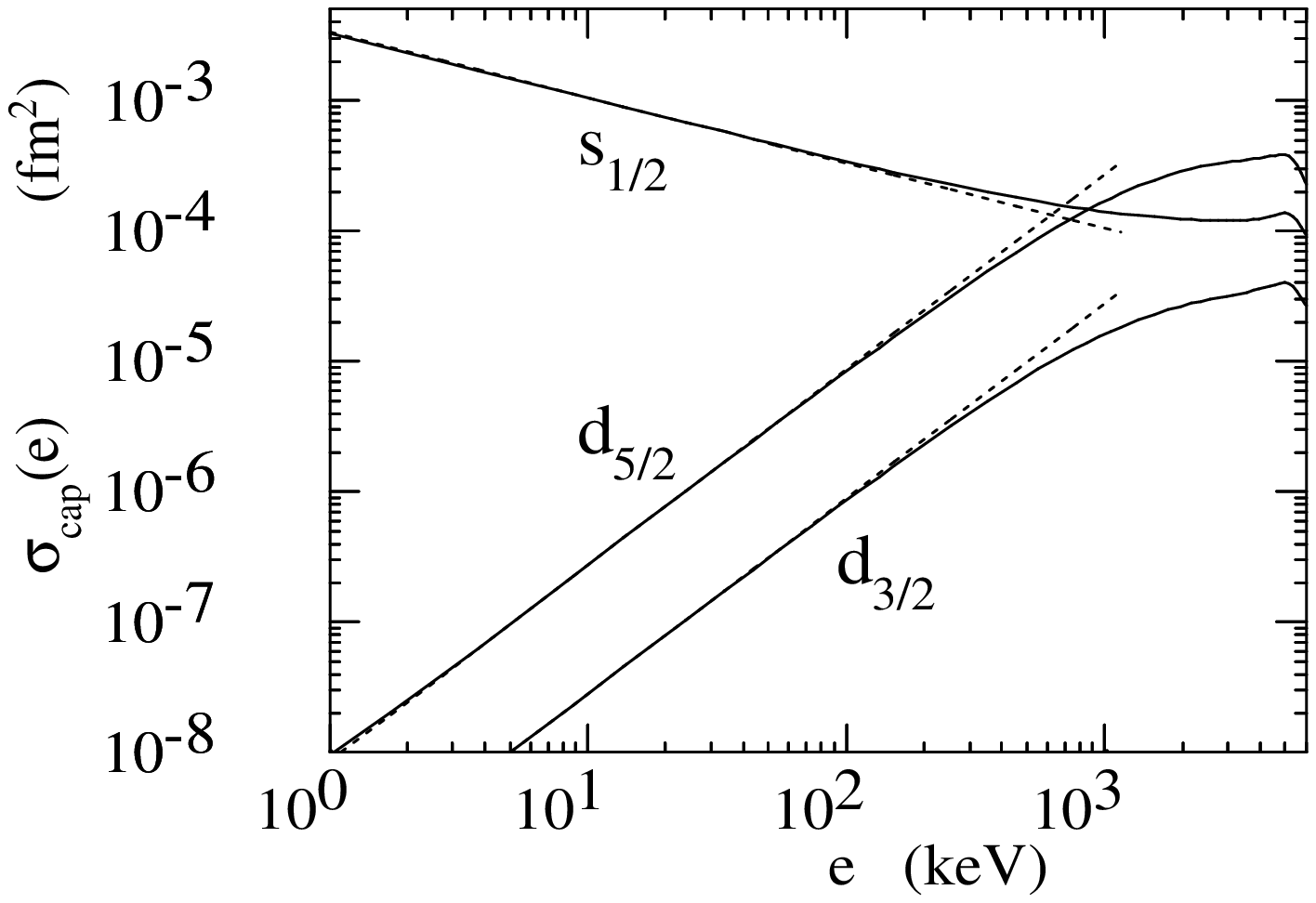}
\caption{
Neutron capture cross sections for three different entrance channels,
consisting of
the ground state with the $3p_{3/2}$ configuration of $^{141}$Sn and an incident neutron
in the partial waves $s_{1/2}$, $d_{3/2}$ and $d_{5/2}$, populating $1^{-}$ states decaying
to the ground state of $^{142}$Sn. The horizontal axis is 
the neutron kinetic energy $e$.  
}\label{fig:n-cap}
\end{figure}

Finally we show in Fig.\ref{fig:n-cap} direct neutron capture cross section 
for $^{141}$Sn in the ground state having  
 the one-quasiparticle 
configuration $3p_{3/2}$ and with the 
E1 decay populating the $0^+$ ground state of $^{142}$Sn.
It is calculated  for  neutron kinetic  energies from
$e=1$ keV to 8 MeV using
Eq.(\ref{ncapture}) and the partial photo-absorption cross sections shown in Fig.\ref{fig:photo-abs-2}.
 We see that
the s-wave capture is dominant at low energies $e\lesim 1$ MeV as expected. It is also
seen that the energy dependence at very low energies $e\lesim 100$ keV obeys 
the power-low scaling  $\sigma \propto e^{l-1/2}$ (with $l$ being the orbital angular momentum of
the partial wave). This threshold behavior arises from the low-energy asymptotics of the neutron continuum
quasiparticle states in the $s_{1/2}$, $d_{3/2}$ and $s_{3/2}$ waves. 
Note, however, that
their absolute magnitudes as well as behaviors at higher energies
differ from the simple single-particle model, as we discussed above.  
The threshold scaling behavior would have been affected if the $s$- and $d$-wave neutron
had low-energy resonance or virtual state in $e\lesim 100$ keV, or if the QRPA correlation 
would have exhibited narrow
resonances in this energy region. Such situation is not seen in the present
example.

\section{Conclusions}

The quasiparticle random phase approximation (QRPA) combined with
the Hartree-Fock-Bogoliubov mean-field model or the nuclear density functional theory
is one of the most powerful frameworks to describe the electro-magnetic responses and the
photo-absorption reaction of neutron-rich nuclei.
In this paper, we have extended  this framework to describe the
{\it direct} radiative neutron-capture reaction of neuron-rich nuclei, 
one of key reactions in the
astrophysical rapid neutron-capture  process. This approach enables one, for the first time,
to take into account
the pairing correlation and the RPA correlations in calculating the direct neutron capture
cross section.

We have formulated a method to calculate {\it partial}  photo-absorption
cross sections corresponding to individual channels of one- and two-nucleon emission decays.   
It is a generalization of the method of Zangwill and Soven, originally  formulated in the
continuum RPA for unpaired systems, to the case of  the
continuum QRPA suitable for pair correlated nuclei.  We  select one-neutron emission channels
in which the decay populates the daughter nucleus in its ground state. We then use
the reciprocity theorem to transform the partial photo-absorption cross section 
to the radiative neutron capture cross section. With improved numerical procedure, we
made it possible to evaluate the neutron capture cross section at very low neutron
kinetic energies of  O(1 keV) and for nuclei with small neutron separation energies.
The theory also enables us to evaluate the branching ratio of the one- and two-neutron
emission decays of the photo-excited states.

Performing numerical calculations for
the photo-absorption of $^{142}$Sn and the neutron-capture
of  $^{141}$Sn, we have shown that  the pairing and the RPA correlations influence
 the results significantly.  It is shown also that the threshold behavior of the
cross sections, governed by the partial waves of the emitted/incoming neutron,
emerges in the present theory.

We remark that in the present work we have neglected the gamma decays from excited to 
excited states.  For example, we find a low-lying collective $2_1^+$ state 
below the neutron separation energy in the present QRPA calculation,
but possible E1 transition from the $1^-$ state populated by the
neutron capture to the collective $2_1^+$ state is not described in the present formalism.
 It is a future problem
to extend  the formalism to include this kind of transitions.
 We also note that neutron capture of even-$N$ isotopes needs to be described
in a separate way. 

\section*{Acknowledgment}

The author thanks T. Nakatsukasa, K. Ogata  and K. Yabana for useful discussion.
This work is supported by Grant-in-Aid for Scientific Research from Japan
Society for Promotion of Science No. 23540294 and No. 26400268.

\section*{Appendix A}

We shall show a derivation of Eq.(\ref{strfn3}).

We note first
\begin{eqnarray}\label{strfn-app}
S(\hbar\omega)&=&
-{1\over \pi}{\rm Im}\int d\vecr \sum_\alpha 
\bar{v}_{\alpha}^{{\rm ext}}(\vecr,\omega)
\delta\rho_{\alpha}(\vecr,\omega)
  \nonumber \\
&=&
-{1\over \pi}{\rm Im}
\int d\vecr \sum_\alpha 
\left(\bar{v}_\alpha^{{\rm scf}}(\vecr,\omega)
-\sum_\gamma\bar{\kappa}_{\alpha\gamma}(\vecr)
\left(\delta\rho_\gamma(\vecr,\omega)\right)^*\right)
\delta\rho_\alpha(\vecr,\omega)  
\nonumber \\
&=&
-{1\over \pi}{\rm Im}
\int d\vecr \sum_\alpha  \bar{v}_\alpha^{{\rm scf}}(\vecr,\omega)\delta\rho_\alpha(\vecr,\omega)
\nonumber \\
& & + {1\over \pi}{ 1 \over 2i}
\left[ \int d\vecr \sum_{\alpha,\gamma}\bar{\kappa}_{\alpha\gamma}(\vecr)
\left(\delta\rho_\gamma(\vecr,\omega)\right)^*
\delta\rho_\alpha(\vecr,\omega) \right. \nonumber \\
&& \hspace{10mm}
-
\left. \int d\vecr \sum_{\alpha,\gamma}\bar{\kappa}_{\alpha\gamma}^*(\vecr)
\delta\rho_\gamma(\vecr,\omega)
\left(\delta\rho_\alpha(\vecr,\omega)\right)^* \right],
\end{eqnarray}
where
\begin{equation}
\bar{\kappa}_{\alpha\gamma}(\vecr)\equiv \left(\kappa_{\alpha\gamma}(\vecr)\right)^*s_\alpha
= \frac{\del^2 E}{\del \rho^*_\alpha(\vecr)\del\rho^*_\beta(\vecr)} s_\alpha = 
\frac{\del^2 E}{\del \rho_\alpha(\vecr)\del\rho^*_\beta(\vecr)}.
\end{equation}
Using the symmetry
\begin{equation}\label{kappa-condition}
\bar{\kappa}_{\alpha\beta}(\vecr)^* =\bar{\kappa}_{\beta\alpha}(\vecr),
\end{equation}
we find that the term in the parenthesis in the last expression
in  Eq.(\ref{strfn-app}) vanishes.
We note also that the linear response equation (\ref{respeq1}) is written as
\begin{equation}
\delta\rho_\alpha(\vecr, \omega)=\int d\vecr' \sum_\beta
R_0^{\alpha\beta}(\vecr,\vecr',\omega)v^{{\rm scf}}_\beta(\vecr,\omega).
\end{equation}
Inserting this into Eq.(\ref{strfn-app}), we obtain Eq.(\ref{strfn3}).

\section*{Appendix B}

In this appendix, we discuss spectral property of $S_d(\hbar\omega)$ in Eq.(\ref{strfn-d}):
\begin{equation}
S_d(\hbar\omega) = -\frac{1}{\pi}{\rm Im}
\sum\sum_{n>m} 
\left\{ 
 \frac{|\bra{nm}V_{{\rm scf}}(\omega)\ket{0}|^2}{\hbar\omega -E_n-E_m+ i\eps}
-
 \frac{|\bra{0}V_{{\rm scf}}(\omega)  \ket{nm}|^2}{\hbar\omega +E_n+E_m+ i\eps}
\right\} .
\end{equation}
It is tempting to expect delta function peaks at $\hbar\omega=\pm(E_n+E_m)$, the
energies of the two-quasiparticle states consisting of bound quasiparticle states
$m$ and $n$, but
this is not the case.

To show this, we return to the linear fluctuation in the state vector 
$\ket{\delta\Phi(t)}= e^{-i\omega t} \ket{\delta\Phi(\omega)}+
e^{i\omega t} \ket{\delta\Phi(-\omega)}$
obeying Eq.(\ref{fluc-state}), which reads in the frequency domain
\begin{equation}
\left( \hbar\omega - \hat{h}_0 + i\eps \right) \ket{\delta\Phi(\omega+i\eps')} 
 = V_{{\rm scf}}(\omega+i\eps')\ket{\Phi_0}.
\label{delPhi1}
\end{equation}
The strength function $S_d(\hbar\omega)$ is then written 
 as
\begin{equation}
S_d(\hbar\omega) =
\sum\sum_{n>m} 
 \frac{\eps}{\pi}|\left\langle {nm} | \delta\Phi(\omega+i\eps')\right\rangle |^2
- \frac{\eps}{\pi} |\left\langle \delta\Phi(-\omega-i\eps')|  {nm}  \right\rangle |^2 .
\end{equation}

We remark here that all the quantities related to the linear response, e.g.,
 $\ket{\delta\Phi(\omega)}$ and  $\delta\rho_\alpha(\omega)$ inherit the
spectral property of the linear response equation (\ref{respeq1}), which
exhibits the QRPA eigen modes. Therefore, in the
discrete energy region $\hbar\omega <S_1 $, the matrix elements
$\left\langle {nm} | \delta\Phi(\omega)\right\rangle$
and $\left\langle \delta\Phi(-\omega)|  {nm}  \right\rangle$
 have poles $\propto 1/(\hbar\omega \mp \hbar\omega_k + i\eps)$,
and hence $S_d(\hbar\omega)$ displays
 delta function peaks at the discrete QRPA eigen energies $\pm \hbar\omega_k$
in the limit $\eps \rightarrow 0$:
\begin{equation}
S_d(\hbar\omega) =\sum_{k,\hbar\omega_k<S_1}  s^d_k\delta ( \hbar\omega - \hbar\omega_k)
- s^d_k\delta ( \hbar\omega + \hbar\omega_k).
\end{equation}
On the other hand, in the continuum energy region $\hbar\omega > S_{1}$,
the matrix elements $\left\langle {nm} | \delta\Phi(\omega+i\eps')\right\rangle$
and $\left\langle \delta\Phi(-\omega-i\eps)|  {nm}  \right\rangle$ are
continuous functions of real $\omega$. Therefore $S_d(\hbar\omega)$ is proportional to
$\eps$ for sufficiently small $\eps$, and it vanishes in the limit $\eps \rightarrow 0$
and for $\hbar\omega > S_{1}$.

\end{document}